# Structural, dielectric and magnetic properties of multiferroic (1-x) La$_{0.5}$Ca$_{0.5}$MnO$_3$-(x) BaTi$_{0.8}$Sn$_{0.2}$O$_3$ laminated composites

S. Ben Moumen, Y. Gagou, S. Belkhadir, D. Mezzane, M. Amjoud, B. Rožič, L. Hajji, Z. Kutnjak, Z. Jaglicic, M. Jagodic, M. El Marssi, Y. Kopelevich and Igor A. Luk'yanchuk.

*Abstract*— High performance lead-free multiferroic composites are desired to replace the lead-based ceramics in multifunctional devices applications. Laminated compounds were prepared from ferroelectric and ferromagnetic materials. In this work, we present laminated ceramics compound by considering the ferromagnetic La$_{0.5}$Ca$_{0.5}$MnO$_3$ (LCMO) and the ferroelectric BaTi$_{0.8}$Sn$_{0.2}$O$_3$ (BTSO) in two different proportions. Compounds (1-x) LCMO-(x) BTSO with x=1 and 0 (pure materials) were synthesized by the sol gel method and x=0.7 and 0.5 (laminated) compounds were elaborated by welding appropriate mass ratios of each pure material by using the silver paste technique. Structural, dielectric, ferroelectric, microstructure and magnetic characterization were conducted on these samples. X-ray scattering results showed pure perovskite phases confirming the successful formation of both LCMO and BTSO. SEM images evidenced the laminated structure and good quality of the interfaces. The laminated composite materials have demonstrated a multiferroic behavior characterized by the ferroelectric and the ferromagnetic hysteresis loops. Furthermore, the enhancement of the dielectric constant in the laminated composite samples is mainly attributed to the Maxwell-Wagner polarization.

*Index Terms*— ferroelectric, ferromagnetic, laminated composites, multiferroics.

## I. INTRODUCTION

NOWADAYS, the development of the electronic technology has become increasingly important. The challenge for researchers is the miniaturization and the multifunctionality involved in many multifunctional devices such as memory devices, transducers, opticals, actuators and sensors [1]–[3]. Among these multifunctional materials, we distinguish the multiferroic compounds consisting of at least two or more ferroic orders such as ferromagnetism, ferroelectricity, ferroelasticity, or ferrotoroidicity [4]. In some multiferroics, strong interaction between the ferroic orders can generate additional functionalities, such as the coupling effect, namely the magneto-electric (ME) effect. In composites containing magnetostrictive and piezoelectric materials, the mechanism behind the ME coupling can be explained as follows. When an electrical field is applied to the composite, the piezoelectric material is strained. This induced strain is transferred as stress to the magnetostrictive material which generates a change in the magnetic moment in the material. The converse effect is also possible, in general, the ME effect in composite materials could be achieved either by the application of an electric field inducing the change in magnetic permeability or by the application of a magnetic field which generates an electric field [5], [6]. Up to now, much work has been done on synthesis of multiferroic materials with high coupling coefficients, composed of single phase multiferroics or composite multiferroics with different connectivity schemes [7], [8]. In contrast to the single phase multiferroics with weak coupling between the ferroic orders which limits their use in multifunctional devices [9], heterostructure composites like 2-2 type, called the laminated composites and grown by alternating ferromagnetic and ferroelectric layers possess generally lower leakage current. This makes the electric poling possible, thus enhancing the (ME) effect [10], [11]. Several processing methods have been developed for the preparation of 2-2 type composites and among them the most often listed are, the tape casting [12], [13], the co-sintering [14], the co-firing [15], and the sintering and bonding using epoxy [16]. In view of the concern with environmental pollution and human health, it is necessary to find appropriate alternative materials for lead-free applications. The calcium-doped lanthanum manganite perovskite (La$_{1-x}$Ca$_x$MnO$_3$) has attracted considerable attention exhibiting a rich phase diagram as a function of the concentration and temperature. It is known by its competition between the super-exchange interactions, which favors antiparallel alignment of neighboring Mn and double exchange interaction, which favors parallel alignment compared with the common ferrites. In particular La$_{0.5}$Ca$_{0.5}$MnO$_3$ (LCMO) would be an attractive candidate as a ferromagnetic component for multiferoic materials since its transition temperature near the room temperature has a beneficial role in the ferromagnetic layers applications. On the other hand tin doped barium titanate BaTi$_{1-x}$Sn$_x$O$_3$ materials have received particular attention due to their important property of tuning the temperature transition with an optimum

S. Ben Moumen, S. Belkhadir, D. Mezzane, M. Amjoud, L. Hajji are with the Laboratory of Condensed Matter and Nanostructures (LMCN), Cadi-Ayyad University, Faculty of Sciences and Technology, Departement of Applied Physics, Marrakech, Morocco. (correspondence e-mail: said.benmoumen@yahoo.fr).

Y. Gagou, M. El Marssi and Igor A. Luk'yanchuk are with the Laboratory of Physics of Condensed Matter (LPMC) University of Picardie Jules Verne, Amiens, France.

B. Rožič and Z. Kutnjak are with Jozef Stefan Institute, Jamova cesta 39, 1000 Ljubljana, Slovenia.

Z. Jaglicic and M. Jagodic are with the Institute of Mathematics, Physics and Mechanics, Jadranska 19, 1000 Ljubljana, Slovenia.

Y. Kopelevich is with the Instituto de Fisica Gleb Wataghin, Universidade Estadual de Campinas, UNICAMP 13083-859, Campinas, Sao Paulo, Brazil doping of Sn at Ti site of BTO. The $BaTi_{0.8}Sn_{0.2}O_3$ (BTSO) was chosen to play the role of the ferroelectric layer because of its transition temperature near the room temperature which coincides with the magnetic transition temperature [17]. In the present study, a 2-2 type bi-layer laminated composite was elaborated by welding a ferromagnetic LCMO and a ferroelectric BTSO ceramics together using the silver paste technique. The structural, dielectric and magnetic characteristics of the multiferroic (1-x) LCMO-(x) BTSO laminated composites were investigated.

## II. EXPERIMENTAL

### II.1. Material preparation

All chemicals were used as purchased without further purification. Both $La_{0.5}Ca_{0.5}MnO_3$ (LCMO) and $BaTi_{0.8}Sn_{0.2}O_3$ (BTSO) ceramic powders with the required compositions were synthesized by sol-gel technique. For this purpose, initially the ferromagnetic LCMO was synthesized, a stoichiometric amount of highly pure nitrates reagents $La(NO_3)_3.6H_2O$, $Ca(NO_3)_2.4H_2O$ and $Mn(NO_3)_2.4H_2O$ were weighed in stoichiometric amounts and dissolved in distilled water, a mixture of citric acid ($C_6H_8O_7$) and tartaric acid ($C_4H_6O_6$) were used as chelating agents and added to the solution. The whole mixture was stirred at 90 °C for about 7 hours. By removing the solvent the solution turned into yellow golden gel. Finally the as prepared powder was heat-treated at 850°C for 8h in air with a heating rate of 2 °C min$^{-1}$ in order to get LCMO nanopowder. The ferroelectric ceramic (BTSO) was synthesized by dissolving in acetic acid a precisely weighted amounts of barium nitrate ($Ba(NO_3)_2$) and tin chloride ($SnCl_4$), after their total dissolution titanium isopropoxide ($C_{12}H_{28}O_4Ti$) was added. A milky solution was obtained in which ammonium hydroxide ($NH_4OH$) was dropwise added while heating at 80°C for about 1h till getting a transparent solution that turned into a gel by removing the solvents. The resulting product was then calcined at 1000 °C for 2 h. Lastly; the obtained LCMO and BTSO powders were uni-axially pressed into pellets and sintered at 1000°C/2h and 1300°C/2h, respectively. The laminated (1-x) LCMO-(x) BTSO composites ceramics were bonded by using a silver conductive paste purchased from Aldrich. The paste was applied on the surfaces of the two ceramics and then the laminated ceramics were co-sintered at 1000 °C for 30 min.

### II.2. Characterization

X-ray diffraction patterns were recorded at room temperature on a diffractometer X'Pert PRO operating with geometry θ-θ using Cu-Kα radiation in the range of 10-80°. The Rietveld analysis of the X-ray diffraction patterns was carried out for both LCMO and BTSO samples. Scanning electron microscopy images and EDX spectra were recorded using a scanning electron microscope (Tescan Vega 3 SEM). For the dielectric measurements the ceramics were coated with silver electrodes on top and bottom surfaces. The temperature and frequency dependent spectra (173 to 353K; 100 Hz to 1 MHz) were obtained by using a Solartron SI 1260 Impedance Analyzer. Magnetic measurements were performed on a Physical Property Measurement System (Quantum Design, PPMS- DynaCool) apparatus operating in temperature range 2-300K and Magnetic field in the range 0 - 9T. Magnetization was measured using VSM (vibrating sample magnetometer) method which is integrated in this system. The P-E measurements were carried out by using a TF Analyzer 3000 designed by AixACCT.

## III. RESULTS AND DISCUSSION

### III.1. Chemical analysis

The oxygen stoichiometry of the ferromagnetic LCMO sample was evaluated using the oxidation reduction method as described by Yang et al [18]. The value of 3.0066 was found almost equal to the theoretical value of 3. Furthermore the $Mn^{3+}$ and $Mn^{4+}$ ions percentage is checked quantitatively. The chemical analysis has been done as follows: LCMO powder was dissolved in a mixture solution of oxalic acid dehydrate $H_2C_2O_4$ and dilute sulfuric acid $H_2SO_4$ with heating at about 50 °C. The compound reacts with $H_2C_2O_4$ which reduces both $Mn^{3+}$ and $Mn^{4+}$ into $Mn^{2+}$. The excess of $H_2C_2O_4$ solution was thereafter titrated using $KMnO_4$ solution. The experimental results are found to be 48.67% for $Mn^{3+}$ and 51.33% for $Mn^{4+}$. These results agree with theoretical data and confirm the stoichiometry of our sample.

### III.2. X-ray diffraction

Fig.1 (a) and (b) represent the room temperature X-ray refined results for LCMO and BTSO that were obtained by using FullProf software [19]. The patterns clearly indicate that both calcined LCMO and BTSO crystallized in the perovskite structure with no evidence of any impurity phases. As shown in the Rietveld refinement, LCMO is crystallized in the orthorhombic structure with space group *Pnma* (see Fig.1 (a)), while BTSO ceramic crystallized well in the cubic structure with Pm3m space group as suggested in Refs [17], [20] for Sn doped BaTiO3 materials with a tin concentration higher than 10% (see Fig.1 (b)). We have started the profile refinement with the scale, zero point and background parameters followed by the unit cell parameters. Then a pseudo-Voigt function was used to describe the individual line profiles, the peak asymmetry and preferred orientation corrections are applied. Finally, the positional parameters and the individual isotropic parameters are refined. The little difference between calculated and experimental patterns and the values of the fit-quality indicator $\chi^2$ demonstrate a successful refinement. All Rietveld refinements parameters are gathered in Table. I.

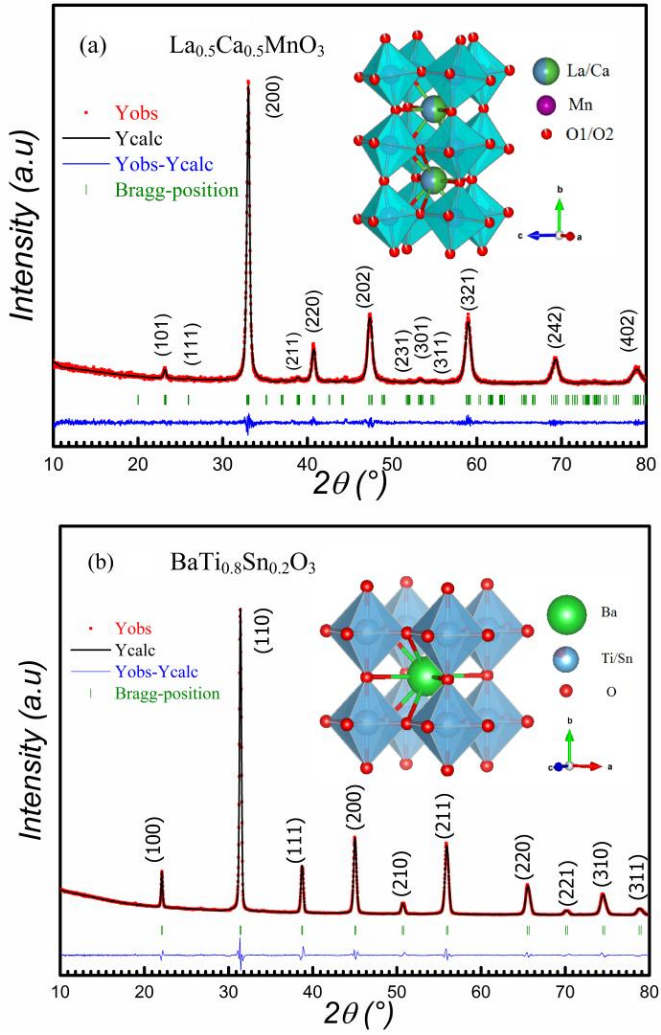

Fig. 1. Room temperature X-ray diffraction pattern at 300 K for (a) $La_{0.5}Ca_{0.5}MnO_3$ and (b) $BaTi_{0.8}Sn_{0.2}O_3$ ceramics including the experimental and calculated profiles as well as their differences. The insets show the structure construction as obtained from refined X-ray results.

TABLE I
STRUCTURAL PARAMETERS OF LCMO AND BTSO AS OBTAINED FROM RIETVELD REFINEMENT.

| Atomic positions | Latice parameters | Reliability factors (%) |
|---|---|---|
| La /Ca (0.0132, 0.2500, -0.0044) | a= 5.4220 Å | Rp=15.6 |
| Mn (0.0000, 0.0000, 0.5000) | b= 7.6467 Å | Rwp=16.8 |
| O1 (-0.0137, 0.2500, 0.4434) | c= 5.4514 Å | $\chi^2$= 0.9 |
| O2 (0.7151, -0.0271, 0.2862) | V= 226.01 Å$^3$ | |
| | | |
| Ba (0.5000, 0.5000, 0.5000) | a=b=c= 4.0264 Å | Rp=10.4 |
| Ti/Sn (0.0000, 0.0000, 0.0000) | V= 65.2742 Å$^3$ | Rwp=10.1 |
| O (0.5000, 0.0000, 0.0000) | | $\chi^2$=4.97 |

III.3 SEM and EDX investigations

Fig. 2(a) and 2(b) display the SEM micrographs and particles distribution of LCMO and BTSO ceramics while Fig. 2 (c) depicts the cross-section of the laminated LCMO/Ag/BTSO composite taken on the cleavage surface. The sintered LCMO phase (Fig. 2 (a)) is found to be characterized by homogenous and spherical shaped grains with uniformly distributed grains ranging in size from 10 to 100 nm and the pores are distributed randomly throughout the sample while the BTSO phase (Fig. 2 (b)) was found to consist of grains of about 2–5 μm in diameter with the formation of aggregates. In order to assign the two distinguishable phases, EDX analysis were carried out near the two interfaces confirming the existence of the characteristic elements of the two phases namely LCMO and BTSO. Besides, no diffusion has been detected as shown in Fig. 2 (d) and (e).

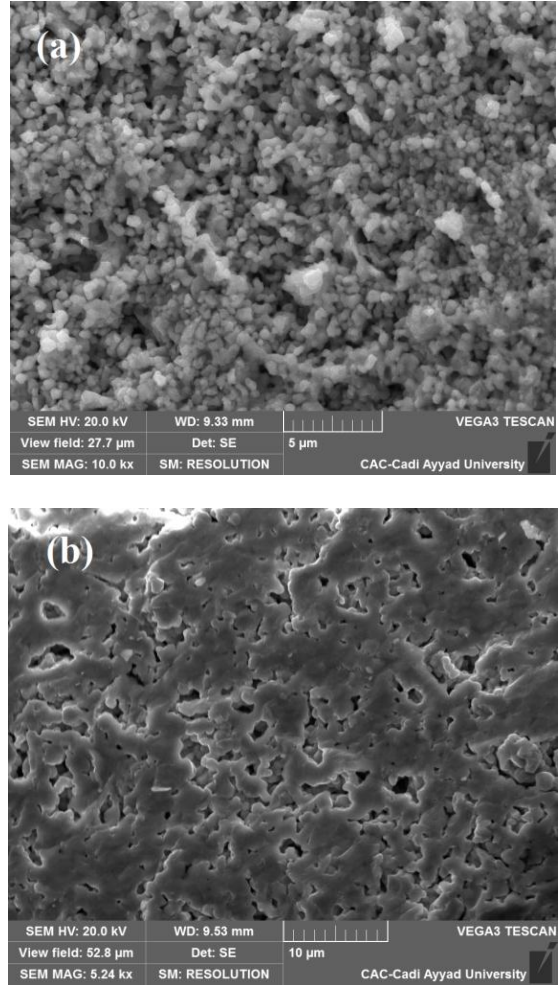

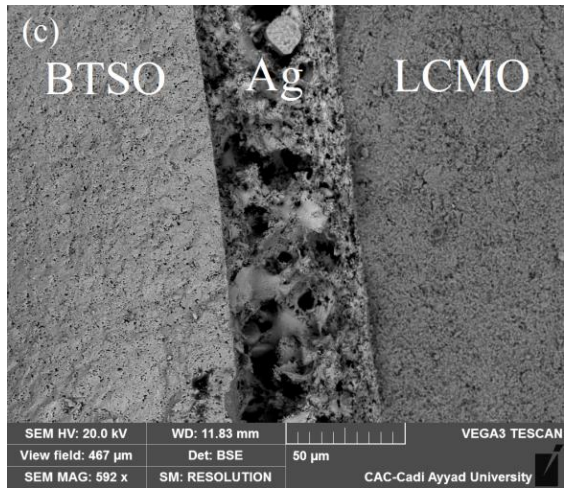

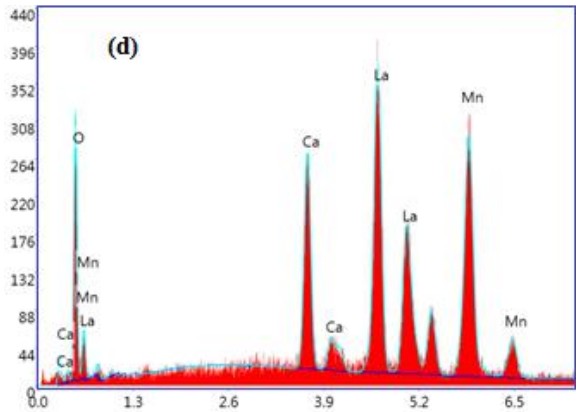

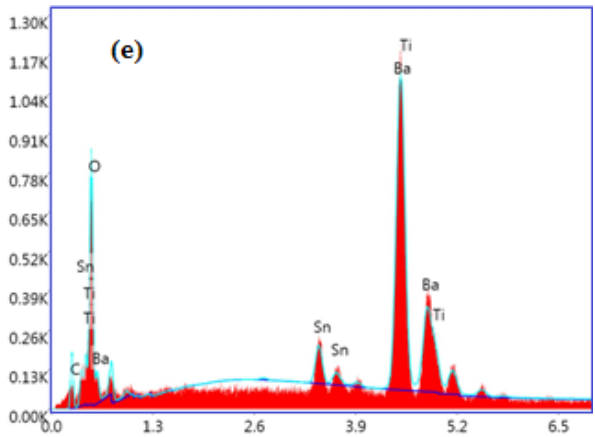

Fig. 2. SEM micrographs of sintered (a) LCMO and (b) BTSO. (c) The cross-sectional image of the laminated LCMO/Ag/BTSO shows regularly defined interfaces, (d) and (e) show the EDX results of the ceramics near the interfaces.

### III.4 Dielectric properties

The temperature dependence of the dielectric constant ($\varepsilon_r$) obtained for the laminated (1-x) LCMO-(x)BTSO samples with (wt%) x=1, 0.7 and 0.5 measured in the temperature range from -100°C to 80°C at different representative frequencies, is illustrated in Fig. 3(a), (b) and (c), respectively. the pure BTSO and 50LCMO-BTSO exhibit a broad ferroelectric-paraeletric phase transition at around Tc=-21°C and a phase transition at Tc=-24°C for 30LCMO-70BTSO sample. The slightly downshifted Curie temperature observed in the laminated 30LCMO-70BTSO composite confirms that the ferromagnetic soldered part doesn't affect much the ferroelectric transition temperature, in consistence with previously reported results [21]. In addition, the values of εr at 1 kHz are found to be 1663, 4153 and 6127 for compounds with x=1, 0.7 and 0.5, respectively. In contrast, the dielectric losses (tan δ) remain invariable with relatively low value at around tan δ=0.002 for all samples. It is interesting to note that permittivity of the laminated composites has been significantly increased compared to that of the pure BSTO as presented in Fig. 3(e). Similar behavior has been reported previously in laminated composites $La_{0.7}Ba_{0.3}MnO_3$–$BaTiO_3$ [22] which could be explained by the Maxwell-Wagner polarization mechanism [23]. Specifically, with an additional contribution in the dielectric polarization, the charge carriers are accumulated at the interfaces of the laminated composites due to the difference in conductivities between the high resistive BSTO and the electrical conductor (magnetostrictive LCMO). Fig. 3(d) shows the hysteresis loops of the laminated composites and the pure ferroelectric sample at T=-30°C. All the laminated composites reveal a clear ferroelectric behavior. Table II summarizes the remnant polarization (Pr) and coercive electric fields (Ec) obtained in the vicinity of the transition temperature at T= -30°C. We note that the values of Pr and Ec decrease while the amount of the ferromagnetic part increases, in agreement with the results of S.B. Li et al [21]. The decrease in Ec value is probably due to the presence of the ferromagnetic LCMO layer. The applied voltage on the ferroelectric part is reduced by passing through the ferromagnetic layer, thus in the laminated composites, the current drop will increase as the thickness of the ferromagnetic layer is increased [24].

TABLE. 2
FERROELECTRIC AND FERROMAGNETIC PROPERTIES OF THE PURE LCMO AND BTSO AND LCMO-BTSO COMPOSITES.

| Sample | Ferroelectric | | Ferromagnetic | | |
|---|---|---|---|---|---|
|  | Pr(μC/cm2) | Ec(Kv/cm) | Ms(emu/g) | Mr(emu/g) | Hc(Oe) |
| BTSO | 0.70 | 1.4088 | - | - | - |
| 30LCMO-70BTSO | 0.38 | 0.3326 | 19.33 | 06.70 | 493 |
| 50LCMO-50BTSO | 0.27 | 0.2995 | 31.13 | 10.02 | 438 |
| LCMO | - | - | 91.35 | 47.12 | 839 |

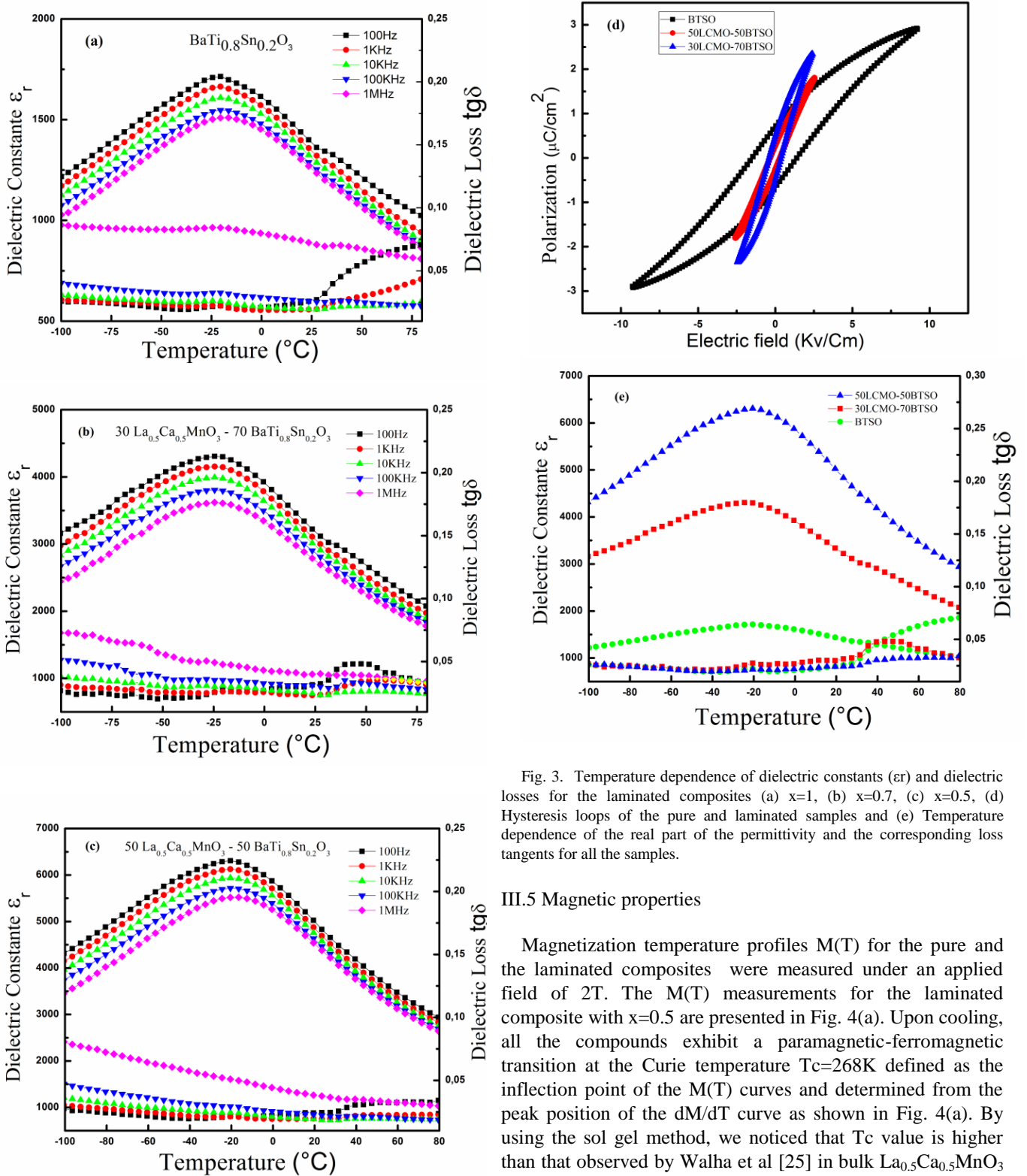

Fig. 3. Temperature dependence of dielectric constants (εr) and dielectric losses for the laminated composites (a) x=1, (b) x=0.7, (c) x=0.5, (d) Hysteresis loops of the pure and laminated samples and (e) Temperature dependence of the real part of the permittivity and the corresponding loss tangents for all the samples.

III.5 Magnetic properties

Magnetization temperature profiles M(T) for the pure and the laminated composites were measured under an applied field of 2T. The M(T) measurements for the laminated composite with x=0.5 are presented in Fig. 4(a). Upon cooling, all the compounds exhibit a paramagnetic-ferromagnetic transition at the Curie temperature Tc=268K defined as the inflection point of the M(T) curves and determined from the peak position of the dM/dT curve as shown in Fig. 4(a). By using the sol gel method, we noticed that Tc value is higher than that observed by Walha et al [25] in bulk $La_{0.5}Ca_{0.5}MnO_3$ (230 K) using the conventional solid state reaction. This result underlines the effect of grain size reduction by using the sol gel technique, which induces a strong ferromagnetic interaction in nanoparticle sample as previously reported by W. Tang et al [26] . The-field dependent magnetization curves of all the laminated (1-x)LCMO-(x)BTSO composites at the maximum applied field of 6 kOe at the temperature of 2K are depicted in Fig. 4(b). All samples show clear hysteresis in accordance with the pure LCMO, thus indicating their ferromagnetic properties. The saturation magnetization values were determined by a linear extrapolation of the magnetization in the high field ranges at 2K. The magnetic properties for various samples are determined and listed in Table II. It is evident that the saturation magnetization (Ms) in the laminated composites increases with increasing LCMO content, as expected, because this parameter depends on the total mass of the magnetic material. In the same way remnant magnetization (Mr) increases with the LCMO quantity while no significant variation in coercive field (Hc) is observed. Such behavior was previously reported in other systems [21], [27]. The noted increase in the values of the saturated magnetization for the composites with the increasing concentration of LCMO indicates the dilution effect of the ferromagnetic phase with non-magnetic one [28]. The evolution of ac magnetic susceptibility in the two samples was very different (Fig. 4(c)). Specifically, in the composite material, the ac magnetic susceptibility appears more smeared and suppressed compared to that of pure LCMO and when approaching the phase transition temperature, it gradually decreases. It appears that the ferroelectric layer has an effect on the magnetic spins reducing their susceptibility and slightly shifting the transition to lower temperatures. Furthermore, to study the (ME) coupling in the composites, we performed the electric field effect on magnetism. Fig. 4(d) represents the M(T) curves measured under different values of the applied electric voltage (0-300 V) across the 50LCMO-50BTSO composite and confirms the absence of an induced shift by the electric field on the M (T) curves or exchange bias effect in a laminated sample. This result underlines the absence of a marked (ME) coupling which can be due to the high conductivity of our samples or to the weak applied electric field [3], [29].

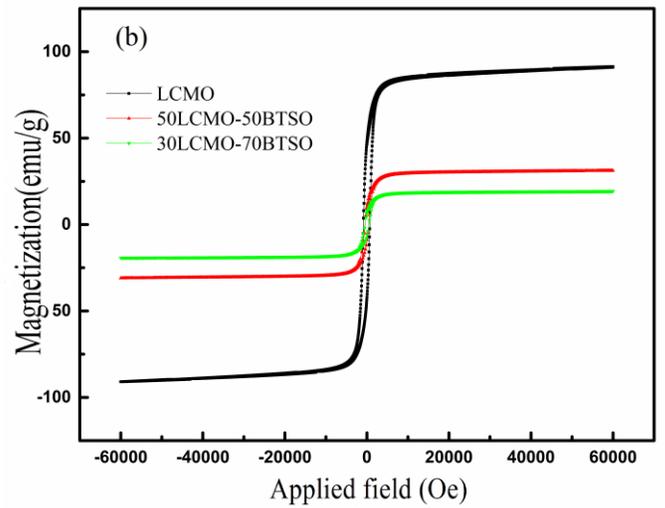

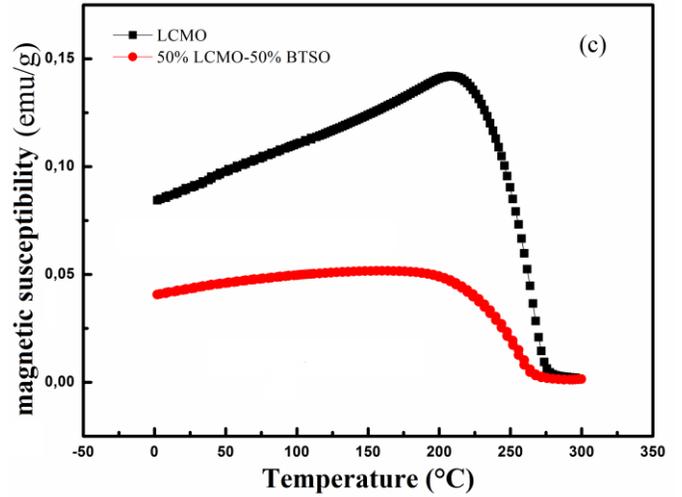

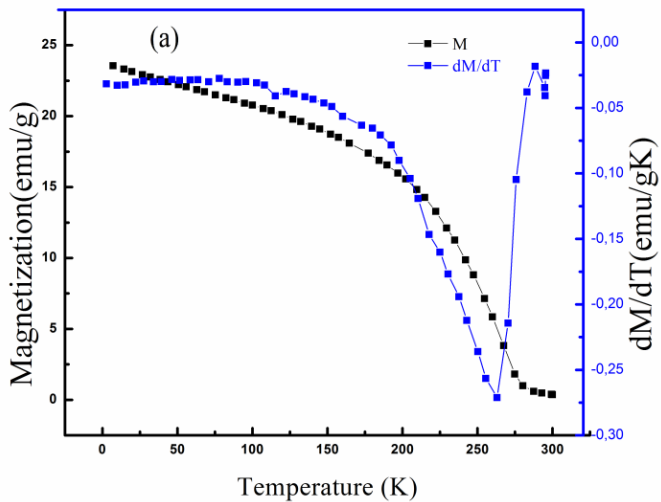

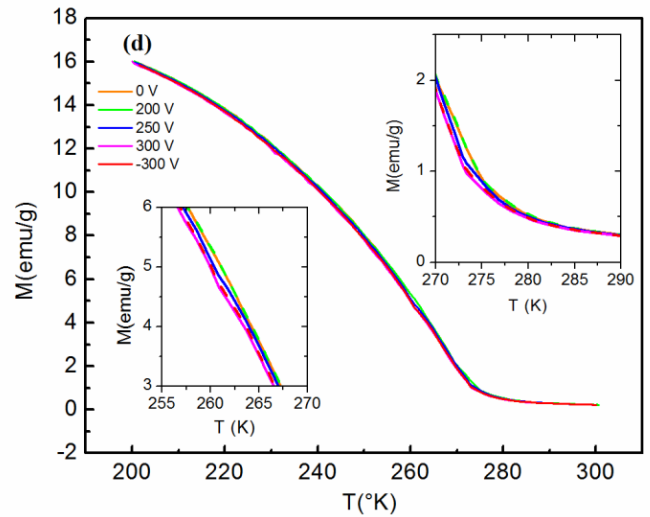

Fig. 4 : (a) Magnetization vs temperature of 50LCMO-50BTSO composite, (b) Hysteresis loops of pure LCMO and the laminated (1-x)LCMO-(x)BTSO composites for x =0.5 and 0.7, (c) ac magnetic susceptibility measured in FC mode for LCMO and 50LCMO-50BTSO composite and (d) Magnetization as function of temperature for the 50LCMO-50BTSO at different electric voltages.

IV. CONCLUSIONS

In this study, lead-free 2-2 type laminated composites (1-x)LCMO-(x)BTSO were synthesized by the sol gel and the silver past techniques. The structural, dielectric and magnetic properties of the composites have been investigated. The SEM and EDS analyses confirm the laminated structure with regular interfaces where no diffusion has been detected. The dielectric constant was greatly enhanced in the laminated composites, a value of ~ 6127 is found in the 50LCMO-50BTSO laminated composite; this εr improvement can be explained by (i) the strong extrinsic interfacial polarization contribution and (ii) the intrinsic effect associated with the charge condensation occurring in systems with cations of mixed valence ($Mn^{3+}$-$Mn^{4+}$). With decreasing the LCMO content the coercive electric fields Ec increases whereas the saturated magnetizations Ms decreases and the Hc coercive fields remains constant. The ferromagnetic performances of the laminated composites were achieved and greatly dependent on the thickness of the ferromagnetic LCMO layer. In order to enhance the (ME) coupling in our samples, other method of soldering will be tried in the future.

ACKNOWLEDGMENT

The authors gratefully acknowledge the financial support of the European H2020-MSCA-RISE-2017-ENGIMA action, Slovenian research agency program P1-0125, the CNRST Priority Program PPR 15/2015, CNPq and FAPESP.